\documentclass{article}

\usepackage{arxiv}

\usepackage[utf8]{inputenc} 
\usepackage[T1]{fontenc}    
\usepackage{hyperref}       
\usepackage{url}            
\usepackage{booktabs}       
\usepackage{amsfonts}       
\usepackage{nicefrac}       
\usepackage{microtype}      
\usepackage{lipsum}
\usepackage{graphicx}
\usepackage{amssymb}
\usepackage{amsmath}
\usepackage{amsfonts}
\usepackage{array}
\usepackage{subfig}
\usepackage{capt-of}
\title{Game theoretic modelling of infectious disease dynamics and intervention methods: a mini-review}

\author{
  Sheryl L. ~Chang \textsuperscript{a,}
  \thanks{Corresponding author} \\
  \textsuperscript{a}Complex Systems Research Group, Faculty of Engineering and IT \\
  The University of Sydney, NSW, 2006, Australia; \\
  \texttt{ sheryl.chang@sydney.edu.au} \\
   \And
Mahendra  Piraveenan \textsuperscript{a,c}\\
\textsuperscript{c}Charles Perkins Centre, The University of Sydney \\
 John Hopkins Drive, Camperdown, NSW, 2006, Australia;\\
  \texttt{mahendrarajah.piraveenan@sydney.edu.au} \\  
   \And
Philippa Pattison \textsuperscript{d} \\
   \textsuperscript{d}The University of Sydney, NSW, 2006, Australia \\
\texttt{philippa.pattison@sydney.edu.au} \\
  \And
Mikhail Prokopenko \textsuperscript{a,b} \\
  \textsuperscript{b}Marie Bashir Institute for Infectious Diseases and Biosecurity \\
   The University of Sydney, Westmead, NSW, 2145, Australia;\\
\texttt{mikhail.prokopenko@sydney.edu.au} \\
}
\begin{document}
\maketitle
\begin{abstract}
We review research papers which use game theory to model the decision making of individuals during an epidemic, attempting to classify the literature and identify the emerging trends in this field. We show that the literature can be classified based on (i) type of population modelling (compartmental or network-based), (ii) frequency of the game (non-iterative or iterative), and (iii) type of strategy adoption (self-evaluation or imitation). We highlight that the choice of model depends on many factors such as the type of immunity the disease confers, the type of immunity the vaccine confers, and size of population and level of mixing therein. We show that while early studies used compartmental modelling with self-evaluation based strategy adoption, the recent trend is to use network-based modelling with imitation-based strategy adoption.  Our review indicates that game theory continues to be an effective tool to model intervention (vaccination or social distancing) decision-making by individuals.
\end{abstract}


\section{Introduction}
Epidemiology plays an essential role in identifying and mapping emerging diseases, assisting health authorities and policy makers in allocating resources for preventive measures or interventions. The use of epidemiological evidence to shape health policy, however, often involves a frustrating delay as the acquisition of the relevant data is time-consuming \cite{who}. To address this challenge, computational modelling of epidemics has emerged as a tool to predict and epidemic dynamics. Yet, disease spread is not the only factor that influences overall epidemic dynamics. Effective interventions, such as vaccination and social distancing, also play a role in curbing epidemic severity. Individuals in affected population groups independently decide whether or not to participate in an intervention. This decision-making process needs to be modelled separately, although it may be interrelated with the overall epidemic spread. 
Individuals in a large population have often diverse interactions with other individuals, in terms of interaction time, number of contacts, nature of interactions etc. This diversity is typically modelled with complex networks aimed to accurately capture the nuances of the interaction patterns. For large-scale epidemics, however, the number of  individuals in the modelled population can easily reach millions, demanding  high computational efforts and lengthy simulation time \cite{eubank2004modelling, Longini2005,GermannKadauEtAl2006,AceMod,Zachreson2018}. This raises the  question of how to construct a rigorous and computationally efficient approach to incorporate interaction patterns into epidemic modelling while modelling the reasoning processes that individuals use to make decisions about intervention measures. The question calls for  innovative modelling methodologies that can successfully and coherently accommodate the three elements  of modelling: infectious disease dynamics spreading through the population, interaction dynamics within the population, and the decision making of individuals.

Game theory studies strategic decision making among
self-interested entities, in various contexts ranging from economics \cite{net_rev2}, to evolutionary biology, computer science, political science, and even quantum mechanics \cite{evo_game,game_von, game_book,game_japan, game_book}. It is often used as a predictive tool to optimize the decision
making in complex scenarios \cite{games_thomas}. Game theory is also well suited to model the decision making of individuals facing a number of intervention options, such as vaccination and social distancing,  to protect themselves from an epidemic spread.  This is reflected  in recent research adopting game theory to model decision making of individuals during contagions. However, to our knowledge there exists no comprehensive review  summarizing the state of the art in game-theoretic modelling of epidemic dynamics and intervention. This motivates our attempt  to review, classify, and identify the current trends in this research field.
\section{Methods}
A three step process was used for selecting papers of interest, confirming to the `Preferred Reporting Items for Systematic Reviews and Meta-Analyses' (PRISMA) \cite{prisma} checklist. As the first step, keywords were used in Scopus to conduct the preliminary search, while the publication year was not restricted. The key words or key word groupings used are:
\begin{itemize}
	\item (epidemi* OR disease* OR infect* OR outbreak OR endemic) AND (model* OR simu* OR dynamics)
	\item (game OR behavio* OR strateg* OR deci*) AND (vaccin* OR social OR preven*)
	\item (networks OR node OR contact OR spatial) 
\end{itemize} 

The second step, the screening process, used the following criteria to narrow down the volume of papers by scanning through titles and abstracts, leaving only those of high relevance for closer inspection.

\begin{itemize}
	\item  \textit {Model}: records should contain a mathematical model that captures the epidemic dynamics, and/or the behaviour change induced after some decision making.
	\item  \textit {Infectious disease}: the modelled disease is able to spread in the population from  person to person through direct contact in any form.
	\item  \textit {Human behaviour}: only individual human behaviour is of interest, and therefore studies of behaviours by institutions (e.g., government) and/or  other life forms (e.g., animals, plants) were removed.
	\item  \textit {Game-theoretical framework}: the decision-making process is modelled under a game-theoretical framework by comparing payoffs of different strategies while assuming that the vast majority of the population consists of rational decision-makers.
	\item  \textit {Preventive measure}: the change in human behaviour is restricted to taking up a preventive measure that may influence the aggregate epidemic dynamics, including vaccination, social distancing, and/or others. However, the decision is purely voluntary and individuals have the freedom to choose something else (i.e., not taking up any preventive measure). 
	\item  \textit {English language}: only records written in English were reviewed.
\end{itemize}

The third step involved reviewing full texts of the filtered articles to confirm their eligibility for inclusion in this review, and  categorizing them. \\

A flow chart in Appendix (i.e.section \ref{PRISMA})  shows the process in more detail. \\
For categorization, we mainly considered two perspectives: categorization in terms of population modelling, and categorization in terms of game theoretic modelling. The first perspective, \emph{population modelling},  can be elaborated as follows:

\begin{itemize}
	\item \textit {compartmental vs network-based}: population models used are either generic, where average traits of various compartments are considered, or network-based, where each individual is modelled separately and the topology of  their interactions is explicitly considered,
	\item \textit {deterministic vs stochastic}: compartmental epidemic models used are either deterministic (using a set of differential equations), or stochastic (breaking the health status to finer state spaces computed in discrete time),
\end{itemize}

Similarly, the second perspective, \emph{game theoretic modelling}, can be elaborated as follows:

\begin{itemize}
	\item \textit {game setting}:  the frequency of the game (i.e whether the game is played iteratively  or non-iteratively within a given epidemic season),
	\item \textit {coupling mechanism}: the coupling mechanism between the epidemic model and the game theoretic model may involve self-evaluation of each individual player based on their past pay-offs, or imitation of other (relatively successful) players.
	\item \textit {sampling mechanism}:  imitation-based coupling can be based on sampling and imitating the population, or sampling individual neighbours.
\end{itemize}

Further contexts for classification are possible: distinction between the intervention methods employed by players as voluntary vaccination or social distancing. However, in the majority of reviewed studies vaccination was used as the intervention mechanism, and so we did not use this context for categorization. Similarly, it is also possible to classify papers based on the pay-off construction methods (i.e., how the pay-off is defined: in terms of monetary value, risk assessment, years to live etc.). Adding this classification, however, would have made the categorization  too convoluted.

\section{Results}
We begin with generic reflections on the existing literature, followed by a detailed analysis and classification.
\subsection{Population modelling: compartmental vs network-based}
The spread of epidemic within a population is commonly represented by compartmentalization, in which each individual is labelled according to their health status. Epidemic dynamics are captured by tracking how individuals move across these compartments. A commonly used compartmentalization approach is known as `S(susceptible),  I(infected), R(recovered)' or SIR \cite{math_epi}. Assuming that the population is conserved, initially the individuals are marked as susceptible to infection. When a disease breaks out, the susceptible individuals encounter those who are infected and, with some probability (i.e., at some rate) are moved to the infected compartment. The infected individuals are eventually removed when they recover (with or without immunity). 

While compartmental models (both deterministic  and stochastic models) assume homogeneous mixing, network models introduce heterogeneous mixing. In homogeneous mixing, which can be assumed only if the size of the concerned population  is large enough \cite{intro_net}, each individual is treated as the average member of the entire population at a given disease stage. Thus, the number of contacts is assumed to be equal  for all individuals at a given disease stage. Conversely, if the the number of contacts at a given point in time varies from one individual to the other, the population is then said to be `heterogeneous'. One common approach to mimicking the varying number of contacts in real world is to model the heterogeneous population as a network, e.g., Erd\"os-R\'enyi random networks \cite{17}, scale-free networks \cite{scale_free}, small-world networks  \cite{small_world}, and other topologies \cite{Pastor-Satorras}. In general, network models are considered a more realistic and flexible approach since they capture the different interactions individuals have in populations \cite{Keeling}. The difficulty lies, of course, in constructing a suitable network accounting for connections that each individual may have within the population. The use of  (complex) networks to represent disease spread is on the rise because the networks  capture both the transmission dynamics \cite{network_game_b2, network_game_bauch}, and the information spread within a contact network with respect to preventive measures \cite{inf_review}, in conjunction with a deterministic or stochastic model that describes epidemic dynamics \cite{sto_network_2010,sto_network_2012,sto_network_2012_memory,sto_network_2013}.

In a contact network, each individual is represented by a node and the contacts with other individuals are represented by edges. The number of edges attached to a node is defined as the node degree and the distribution of the number of degrees for the entire network is called degree distribution \cite{SIR_network}. The degree distribution is shaped by the  network's topology. Different types of networks reflect the inherent differences in social systems \cite{pattison1,Keeling}. 

Networks can remain \textit{static} during an epidemic,   implying that the network degree distribution  is unchanging during the contagion. This is only possible when the epidemic spread is much faster compared to the evolution of contacts among the affected individuals.  On the other hand, networks can continue growing at a rate which is comparable with the rate of epidemic spread ---  that is, they can be dynamic. This review mainly focuses on static networks.

\subsection{Game theoretic modelling }
\subsubsection{Game theory}
Game theory is used to model scenarios where there are a number of intelligent entities, called `players',  which try to make decisions in the face of uncertainty. The players could represent individuals, groups of people, organizations, or computer programs. Each `player' typically tries to optimize their own profit or benefit,  called `payoff' or `utility' of a player. To do so, each player will take a course of action among available actions, termed `strategies'. Typically, the selection of the best course of action for one player will depend on what the other player or players will decide to do. 
\paragraph{Nash Equilibrium}
Nash equilibrium is one of the pivotal concepts in game theory \cite{Nash_eq}, predicting that in a strategic decision making environment, there exists a set of strategies from which no perfectly rational player would benefit by deviating, where rationality is defined as the tendency to maximize one's own utility. 
Nash equilibrium is defined for both pure and mixed strategies \cite{Nash_eq}. A pure strategy is a strategy that is consistently adopted by a  player  during an iterative game (until a conscious decision is made by that player to switch to another pure strategy). If a mixed strategy is adopted, each  pure strategy available to a player is selected according to a premeditated probability distribution which defines the mixed strategy. Therefore, pure strategies can be thought of as special cases of mixed strategies, where each available strategy is selected with a probability of either 1 or 0 at each round of the iterated game.
\paragraph{Rationality}
Recent studies have adopted a more realistic scenario by including irrationality to mimic those individuals whose decisions are not made completely rationally \cite{beh_imi3, beh_imi2}. Here a  lack of rationality could mean either making decisions for the `public good'  (or letting the `public good' influence their decisions), or, while being totally selfish, arriving at decisions which are not at the best interest of the individual concerned, due to some reason.  Some literature \cite{beh_imi3, beh_imi2} included the term $\tilde{\mu}$ (at a rate $\tilde{\mu} \ll 1$)  as a `rationality parameter' to describe players who randomly switch between strategies, or who rarely change regardless of the payoff balance. This irrationality is modelled as $\tilde{\mu}(1-x)$ and $\tilde{\mu}x$ where $x$ is the fraction of the population that adopts a certain strategy. 
\paragraph{Bounded rationality}
It has been observed that in experimental settings, players deviate substantially from the predictions given by Nash equilibrium \cite{kasthurirathna2015}. One key reason for this deviation is the non-perfect, or bounded rationality of the players, due to possible lack of information available about the strategies adopted by the opponents and their respective payoffs, limitations in cognitive capacity of the players or the limitation of computational time available to make the strategic decision \cite{kasthurirathna2015,kasthurirathna2016}.  Advanced equilibrium models, such as the `Quantal Response Equilibrium (QRE)' have been proposed \cite{QRE} to arrive at equilibrium solutions accounting for bounded rationality of players.
\subsubsection{Game setting in epidemic modelling}
In this review, only two-player games are included but a ``player" can be interpreted as an individual or as a fraction of the population. The game can be a one-off game, or can  be an iterated game. This section attempts to categorize the games used in computational modelling of epidemics, based on the game frequency. 

\paragraph{Categorization based on game frequency}
To investigate the coupling between the epidemic model and the game, it is essential to ascertain when and how the game is played in the model. Games are therefore categorized according to the number of times they are played:

\begin{itemize}
	\item Non-iterative: these are stand-alone games that are only played once during a fixed window of time \cite{sto_network_2010, sto_network_2012, sto_network_2012_memory, sto_network_2013,static_bauch, static_group_ind,minority1, minority2}
	\item Iterative: games that are played iteratively within the simulation of the epidemic model \cite{network_game_bauch, network_game_b2,bauch_2005_imi, imi_bauch_bha, samit_bauch_2010_imi, onofrio_2011_imi_vac} \end{itemize}

Non-iterative games are used in modelling diseases where  vaccination can grant  life-long immunity. Since the protection does not wear off over time, these vaccination games are played once only. They are also applicable in cases where vaccination is only available for a short, fixed window of opportunity per infection season (such as influenza vaccine). In some games, vaccination games are played once per epidemic season prior to a forthcoming breakout. The games are independent of the epidemic transmission during epidemic propagation and are therefore, treated as ``non-iterative" although they could be played more than once within a person's lifespan (i.e., once every epidemic season). 

Iterative games on the other hand are played repeatedly during the epidemic simulation. Individuals are given the option to play the game at any time during the course of an epidemic season and their strategies vary over time as the epidemic progresses. This results in a more intricate linkage between the games and the epidemic model since the input (i.e., risk payoffs) to initiate the games relies on the output from the epidemic model, while the input to the the epidemic model itself depends on the decisions players make.

This section describes how iterative and non-iterative games are constructed in relation to the epidemic model. Earlier studies \cite{static_bauch, static_group_ind} used non-iterative games with a deterministic SIR epidemic model to derive analytical solutions in population games. In more recent studies, non-iterative  games have been extended to simulation-driven and network-based applications through the use of either deterministic SIR and self-evaluation, or stochastic SIR and imitation dynamics \cite{sto_network_2010, sto_network_2012, sto_network_2012_memory, sto_network_2013}. In the case of seasonal epidemics,   there is an imaginary window of time between epidemic season $i$ and $i+1$ during which individuals can change their strategy, which is known as ``decision adjustment".

\begin{figure}[htb!]
	\graphicspath{{./fig/}}
	\includegraphics[width=0.9\columnwidth]{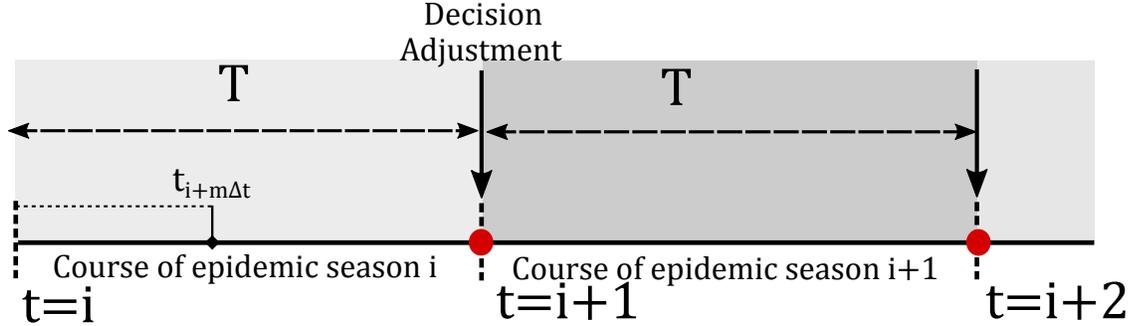}
	\caption{Iterative and non-iterative games in intervention modelling. Non-iterative games are played either prior to the commencement of an epidemic season, or between different epidemic seasons.  Iterative games are played several times during a single epidemic season, and played with a certain time interval $\Delta t$ over the length of a particular season $i$,  and end when that epidemic season $i$ comes to an end. Therefore,  the term `iterative' is used in the context of a particular season, rather than the whole lifetime of an individual.}
	\label{stadyn}
\end{figure}  
Figure \ref{stadyn} highlights the difference between iterative and non-iterative games used in intervention modelling. In  a particular epidemic season $i$, which has a length of $T$, the spread  of infection starts from time $t =  i$ and simulated with time-steps with size   $\Delta t$ until it reaches $t = i+1$.   Iterative games are typically played at  each of the $m$ time steps, where  $m =  \frac{T}{\Delta t}$ over the course of epidemic season $i$.  There is a window of time between epidemic season $i$ and $i+1$ (which, in reality, may be several months) during which individuals can change their strategy.

Iterative games can be played in scenarios where the decision to be made is whether to take prophylactic measures or not, or where the decision to be made is whether to vaccinate or not.  In the first case, the decision is reversible as one can easily change from adopting self-protection to living in a normal lifestyle, or vice versa. Therefore, the strategy can change for every iteration. In the second case, the player may not be able to switch strategy at every iteration, because vaccines typically provide protection over a certain period of time, and the length of protection relies solely on the strength of the vaccine, and may be greater than  $\Delta t$. For re-occurring diseases such as influenza, vaccines normally last until the end of the epidemic season $i$ and wear off before the next wave of epidemic at season $i+1$.  Therefore, a player who plays the vaccination game iteratively within the season is able to vaccinate any time within the season, but cannot undo the decision to vaccinate within that particular season. Hence, in certain iterative vaccination games, one strategy (not to vaccinate) is reversible whereas the other strategy (to vaccinate) is not reversible. Individuals then face the same decision ---  whether to vaccinate or not ---  in epidemic season $i+1$. Some vaccines, however --- typically those for childhood disease --- provide life-long protection so there is no need for re-vaccination for vaccinated individuals.  In such cases also, iterative games can be used to model the behaviour of players, however there will be no `season' involved, and one strategy (to vaccinate) is irreversible life-long. On the other hand, vaccination games can be non-iterative if the vaccine concerned should be administered within a certain `window of opportunity', either once in a lifetime or between seasons, and not during the epidemic propagation.  If no restraint is imposed on the timing of vaccination, iterative games can be played when individuals refer to the current disease prevalence to decide whether they need to vaccinate.

\subsection{Coupling mechanism: decision-making and epidemic dynamics}
To understand the intricate coupling mechanism between the infectious dynamics  model and the games used to model intervention decision making, let us first recall that there are two ways of constructing payoffs: through an individually-based self evaluation, or an imitation influenced by the population's attitude. Although the core concept of payoffs comparison still holds, there is a clear conceptual difference between the two mechanisms: through self-evaluation, individuals rely on their memory and  personal perception of the disease, whilst imitation dynamics puts individuals into an environment where personal decisions are influenced by the choices of the population, or their neighbours, where the neighbours are determined by the topology of the contact network.

The number of games played per season  ultimately determines how games are coupled with the epidemic model. For non-iterative games,  intra-season epidemic dynamics is not dynamically influenced by the games. On the other hand, iterative games are coupled and iterated with the epidemic model at each time step, which suggests that the epidemic dynamics must be traceable at each time step. Stochastic models are often not coupled with iterative games because the stochastic processes make it hard to track the epidemic's influence on individual players with time.  In comparison, deterministic models can be coupled with iterative games \cite{onofrio_2011_imi_vac,bauch_2005_imi,imi_bauch_bha,samit_bauch_2011_delay}. The assumption of homogeneity in deterministic models, however, shifts games to the population level since every individual in the population is treated the same. There are cases where the compartments in deterministic SIR models are further refined by dividing each health compartment into gender or age groups \cite{USA_Israel}. It is impossible, however, for deterministic models to zoom down  to capture the detail of individual differences. Network models position individuals into a constrained environment in which only a finite set of contacts can be reached. This feature naturally focuses on the individuals and is mainly adopted in network-based games. 
Networks can be used to model both disease spread and imitation dynamics sampling neighbours, or imitation dynamics only while a stochastic SIR governs the epidemic dynamics. In the latter case, epidemic dynamics are modelled separately, using stochastic models whose output is entered into the networks \cite{sto_book, sto_network_2010, sto_network_2012, sto_network_2012_memory, sto_network_2013}.  
The following subsection reviews the synthesis between different games using self-evaluation or imitation dynamics and epidemic models. 

\subsubsection{Self-evaluation}
The early epidemic models that incorporated games \cite{static_bauch} considered human responses through self-evaluation, whereby non-communicating and self-centred individuals make decisions purely by themselves after evaluating payoffs from the two strategies. The evaluation process generally takes the current disease prevalence and/or memories of previous events into account. Self-evaluation was used extensively in earlier studies \cite{minority1, minority1} because of its simplicity but has slowly lost prominence due to its inability to capture the influence of human interaction. Nevertheless, self-evaluation provides a foundation for game-theoretic frameworks used in recent studies.  

\paragraph{Non-iterative games with self-evaluation.}
Early studies \cite{static_bauch, static_group_ind} incorporated game theory into epidemic modelling through static population games in which each individual in the population weighs the costs and benefits of two strategies, aiming to maximize individual gain. Voluntary vaccination is the most widely studied concept \cite{static_bauch, static_group_ind, vac_influenza, reluga_galvani}, where the dilemma faced by a player focuses on risks and benefits associated with a defined vaccine. Commonly coupled with a deterministic SIR model, such games take some parameters of payoff functions  from the epidemic model, like the infected fraction of the population for example. After evaluating the risks of the different strategies, individuals make decisions on voluntary vaccination, which can be fed back to the epidemic model as an input updating the vaccination coverage (i.e., the fraction of the vaccinated population). This method has been applied to studies investigating the optimal level of voluntary vaccination coverage for childhood diseases and vaccine scare \cite{static_bauch}, vaccination policy for smallpox in preparation for a possible outbreak \cite{static_group_ind}, H1N1 vaccination strategies \cite{mon_cost2}, influenza vaccination \cite{vac_influenza} and delayed vaccination \cite{samit_bauch_2011_delay}. 

An example of the non-iterative population game in relation to voluntary vaccination \cite{static_bauch} involves a scenario in which each individual in the population of size $N$ decides on the vaccination strategy $S$, represented by the probability of vaccination. The proportion of the vaccinated population is represented by $p$. The payoff evaluation is denoted by morbidity risk $r$;  a relative parameter defined as the ratio between the morbidity risks from vaccination and infection as $r=r_v/ r_i$. The risk of infection at a defined vaccination coverage with vaccination fraction $\theta$, is denoted as $\pi_{\theta}$. The expected payoff of an individual playing strategy $S$ is then given as follows:

\begin{equation}
\begin{aligned}
E(p, \theta)=-rS-\pi_\theta(1-S)
\end{aligned}
\label{vac_pop}
\end{equation}

The aim is to obtain the optimal vaccination strategy $S^*$, known as the \textit{Nash equilibrium}, so that no player can improve his or her payoff by changing to a different strategy. Since the population is homogeneous in terms of contact patterns in population games, the optimal vaccination coverage for the population is equal to the probability of vaccination for each individual, $\theta^*=P^*$. The risk of infection $\pi$ is derived from the epidemic model and the value reduces with greater $\theta$. If $\theta$ exceeds the threshold for herd immunity $\theta^{'}=1/R_0$, $\pi_{\theta \geq \theta^{'}}=0$, the risk of infection becomes zero and the disease is naturally eradicated. To obtain the Nash equilibrium, $S^*$,  the payoff gain to an individual playing $S$ in the population is expressed as \cite{static_bauch}:

\begin{equation}
\begin{aligned}
\Delta E=(\pi_{(N-1)S^*+S}-r)(S-S^*)
\end{aligned}
\end{equation}

A unique strategy $S=S^*$ exists for any given relative risk, $r$, and $\Delta E$ is strictly positive. There are two pure Nash equilibria $S^*=0$ or $1$ and a mixed equilibrium $0<S^*<1$ \cite{static_bauch}. The condition to achieve these equilibria is affected by $r$ and $\pi$. When the vaccine is perceived as more risky than infection $r\geq\pi_0$, the Nash equilibrium is never to vaccinate. If $r<\pi_0$, the Nash equilibrium is to vaccinate with non-zero strategy $S^*$ at vaccination coverage $\theta^*\in (0,\theta^{'})$ such that $\pi_{\theta^*}=r$. Therefore, the mixed Nash equilibrium $S^*$ yields a suboptimal vaccination coverage $\theta^*>\theta^{'}$ \cite{static_bauch}.

Non-communicating individuals often act in their own self-interest and do not share their decision with others \cite{minority1, minority2}. The independent decisions may be made between epidemic seasons (i.e., remain static during a season), as shown in Figure \ref{stadyn}. 
The decision-making process relies on one's experience from the last epidemic season, with an assigned value $V^i_n$ and two independent parameters, memory $s$ and adaptability $\varepsilon$. Taking vaccination games as an example, the probability $\omega_{n+1}^{i}$ that an individual $i$ chooses to vaccinate is given in Equation \ref{min_gam}:

\begin{equation}
\begin{aligned}
\omega_{n+1}^{i}=(1-\varepsilon)\omega_n^i+\frac{\varepsilon V_{n+1}^i}{(s^{n+1}-1)/(s-1)}
\end{aligned}
\label{min_gam}
\end{equation}

with $0<\varepsilon<1$ and $V_{n+1}^i=sV^i_n$. The self-evaluation in relation to the possibility of vaccination is coupled to the epidemic model through  $V^i_n$ which is dependent on the vaccination coverage $p$ and the probability of infection at current vaccination coverage $q(p)$. Some studies \cite{static_group_ind, mon_cost, mon_cost2} use an added compartment in the  deterministic SIR model  representing the `vaccinated' fraction $V(t)$ to explicitly track the vaccinated population.
\paragraph{Iterative games with self-evaluation.}
An individual can play the game at each time step when coupled with a network model in which each individual is seen as a node with contacts as links. In this case, although the health compartmentalization of the population is still used to identify an individual's health status, the population fraction of each health compartment is no longer calculated mathematically. Instead, the epidemic propagation is modelled through a single parameter, the total probability $\lambda$ that a node becomes infected each day with $n_{inf}$ infectious neighbours. $\lambda$, as the output from the epidemic model, is used to construct payoffs; consequently the decision on vaccination affects the number of infected neighbours for a node and the probability of infection. The process is dynamic because it is iterated at each time step (i.e., each day in this case), forming a loop-like feeding pattern \cite{network_game_bauch, network_game_b2}.  An example is given in Equation \ref{net_game}, taken from \cite{network_game_bauch}.
\begin{equation}
\begin{cases}
\lambda_{perc}=1-(1-\beta_{perc})^{n_{inf}} \\
E_{inf}=(1-\lambda_{perc})\alpha+\lambda_{perc}[(1-d_{inf})L] \\
E_{vac}=(1-d_{vac})L \\
\end{cases}
\label{net_game}
\end{equation}
where $\beta_{perc}$ is the perceived probability of node-to-node transmission, $E_{inf}$ is the payoff for non-vaccinators and $E_{vac}$ is the payoff for vaccinators; $\alpha$ is the payoff for individuals with continued susceptibility, $L$ is the payoff for individuals with lifelong immunity, $d_{inf}$ the probability of death due to infection, and $d_{vac}$ is the probability of death due to vaccine-related complications.

\subsubsection{Imitation dynamics}
People learn to adopt new strategies by imitating others in the population who have adopted more successful strategies. This learning process, or imitation dynamics, can take place at a population or individual level. The fundamental concept at both levels is similar: an individual encounters others in the population and  stochastically adopts their strategy if they have received higher payoffs. What differentiates the two levels is the assumption of homogeneity; the population level adoption process assumes that each individual has the same chance of meeting the rest of the population, whereas the individual level adoption process assumes that the contacts of an individual is limited to his/her neighbours in a defined network.
\paragraph{Individuals imitate the population.}
In a homogeneous population, people are divided into two groups according to their strategies. An individual playing with strategy $A$ encounters those playing with strategy $B$ at a constant rate. If there is an expected gain in payoff by switching from strategy A to strategy B, the individual would do so after comparing payoffs. \cite{bauch_2005_imi, imi_bauch_bha, samit_bauch_2010_imi, onofrio_2011_imi_vac}. \\
Assuming the fraction of the population playing strategy A is $x$, the rate of change of $x$ is defined as:
\begin{equation}
\frac{dx}{dt}=\sigma \Delta E x(1-x) \\
\end{equation}
Where  $\Delta E=E(A)-E(B)$, and $E(A)$ and $E(B)$ are the payoffs for playing strategy A  and strategy B respectively, and $\sigma$ is the sampling rate.\\
\begin{figure}[htb!]
	\centering
	\graphicspath{{./fig/}}
	\includegraphics[width=0.9\textwidth]{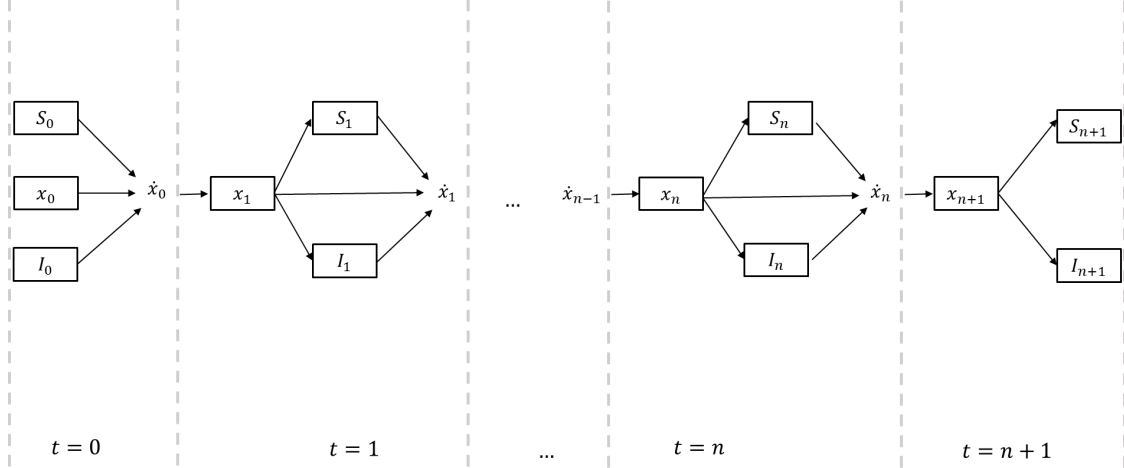}
	\caption{Coupling mechanism for a population game adopting imitation dynamics in a SIR deterministic model. Here, $S_0$ and $I_0$ represent the initial proportions of susceptible and infected people in the whole population respectively. Similarly, $x_0$ represents the initial fraction of vaccinated people in the population, and $\dot{x}_0$ represents the rate of change of this fraction at time $t=0$.  Similarly, $S_n, I_n, x_n$ and $\dot{x}_n$ represent the fraction of susceptible individuals, the fraction of infected individuals, the fraction of vaccinated individuals, and the rate of change of the fraction of vaccinated of individuals, at time $t=n$ respectively.}. 
	\label{Coupling_imi_pop}
\end{figure}
By adopting imitation dynamics at population level, the rate of change for each compartment in the SIR epidemic model depends on the fraction of the population choosing a certain strategy. The coupling runs in the following scheme: with known initial conditions for the number of infected($I_0$ and the number of susceptible $S_0$ from the SIR model and the fraction of the population $x$ playing strategy $A$ in the game, $\frac{dx}{dt}$ is computed and updated $x$ after playing the game by comparing the payoff at each time step;
Typically, with known initial conditions for the number of infected $I_0$ and the number of susceptible $S_0$ from the SIR model, and for the number of vaccinated $x_0$ from the vaccination game, the fraction of the population $x$ playing strategy $A$ in the game, $\frac{dx}{dt}$ is computed and  $x$ is updated after playing the game by comparing the payoffs at each time step. The updated value of $x$ is used, in turn, to update the size of each compartment in the SIR model. If  $E(A)$ and $E(B)$ are dependent on the disease prevalence $\lambda$,  $\frac{dx}{dt}$ continue to depend not only to $x$, but also  $I_t$ and  $S_t$. 
The schematic of this mechanism is shown in Figure \ref{Coupling_imi_pop}. The simulation is iterated until the epidemic spread is eradicated or it reaches an endemic state\cite{bauch_2005_imi, beh_imi1, beh_imi2, beh_imi3}.

An example of how a iterative vaccination game with imitation dynamics sampling the population coupled with a deterministic SIR epidemic model \cite{bauch_2005_imi} follows this scheme:
\begin{equation}
\begin{cases}
\frac{dS}{dt}=\mu(1-x)-\beta SI-\mu S \\
\frac{dI}{dt}=\beta SI -\gamma I-\mu I \\
\frac{dR}{dt}=\gamma I-\mu R+\mu p \\
\frac{dx}{dt}=\sigma x(1-x)[-r_v+r_i mI] \\
\end{cases}
\label{imi_pop}
\end{equation}
\paragraph{Individuals imitate neighbours}
In this adoption, the concept of imitating others' behaviour still holds, but an additional constraint is applied such that individuals can only learn from their neighbours in a topologically defined network. Assuming $i, j$ are connected individuals in a network
$E_i, E_j$ are their respective payoff for the strategy that they currently own. Strategy imitation occurs between epidemic seasons (i.e., during the decision adjustment period shown in Figure \ref{stadyn}) and each individual decides whether to change his/her strategy for the upcoming epidemic season. For an individual $i$, he/she selects a random neighbour $j$ to compare payoffs. Naturally, if $j$ plays with a different strategy and has a higher payoff, $i$ is more likely to imitate $j$'s strategy and thereby change strategy \cite{sto_network_2010, sto_network_2012_memory, sto_network_2012,sto_network_2013, game_japan}. The probability that an individual $i$ adopts an individual $j$'s strategy is given by the Fermi function:
\begin{equation}
f(P_j-P_i)=\frac{1}{1+exp[-\kappa(P_j-P_i)]}
\end{equation}
where $\kappa$ is the strength of selection $0<\kappa<1$ \cite{sto_network_2010}.
Wells et al., 2011 \cite{sto_imi_pop} places individuals on a homogeneously distributed network that models the disease transmission and vaccination strategy adoption. The disease spreads if a susceptible individual has an infected contact (known as the index case). Individuals make decision on vaccination by self-evaluation (comparing the payoff between vaccination and non-vaccination) and imitation (adopting the most prevalent strategy among all contacts). The vaccination decision is finalized on the first day when the infected index case shows symptoms. The game is played continuously until the simulation time is reached. The time window of vaccination is set differently from previously stated papers: there is no fixed time window of vaccination for all individuals to decide on whether to vaccinate or not, as seen in \cite{sto_network_2010, sto_network_2012_memory, network_game_b2, sto_network_2012}; instead, individuals decide on vaccination on the first day when one of their infected contacts symptomatic, at any time step during the epidemic season.  If at that time step they decide not to vaccinate however, they face the same decision when another of their neighbours become symptomatic. Therefore, they have multiple opportunities to decide to vaccinate, though a vaccination decision, of course, is not reversible within that season. Therefore, \cite{sto_imi_pop} use an iterative game though the iteration does not depend on  a fixed time-period.
Feng et al., 2018 \cite{comp_imi_nei} and Li et al., 2017 \cite{net_imi_nei} use non-iterative games which utilize  similar imitation game settings at individual level as seen in \cite{sto_network_2010, sto_network_2012_memory, network_game_b2, sto_network_2012} where individuals imitate the most successful strategy from a randomly chosen neighbour, but combine the imitation game with different population models. While \cite{comp_imi_nei} uses a simple SIR model by dividing the population into three health classes,  \cite{net_imi_nei} divides the population into subclasses by the number of contacts first, then further divides each subclass into three health compartments (S, I, R).  

\subsection{Additional contexts}
\subsubsection{Game context  in epidemic modelling}
In the context of epidemic spread,  individual decision-making is driven by two main types of preemptive measures:
\begin{itemize}
	\item Vaccination: an individual has the choice of vaccinating or not vaccinating in the face of an epidemic.
	
	\item Social distancing: also referred to as behavioural change, this refers mainly to prophylactic actions that reduce the chance of infection.  An individual has the choice to either change their behaviour in order to minimize contact (for example, drive rather than taking the train), or  live normally. \end{itemize} 

In a population, each individual is seen as a player who faces the trade-off between perceived benefits and perceived costs of protection. By adopting protective measures, the probability of infection is reduced but some financial/social cost, or even health risks, is increased. However, one might succeed in avoiding infection  without any cost if they do not take protective measures but every potential contact has taken them. Therefore, an individual's decision is influenced by the behaviour of others,similar to classical scenarios considered in game theory.

In vaccination games, individuals opt to vaccinate with the hope of gaining immunity and avoiding the infection. From a modelling point of view, the core concept of vaccination games is to change individual's health status by moving susceptible individuals into the vaccinated class. Social distancing games, on the other hand, only have the ability to modify infection-related parameters; for example, individuals who adopt protective measures will have a lower transmission rate but are still prone to infection due to remaining in the susceptible class. 
\paragraph{Vaccination}
In vaccination games, three conditions affect the epidemic model and, consequently, the payoff construction. These are: 
\vspace{0.5em}
\begin{itemize}
	\setlength\itemsep{0.25em}
	\item Length and coverage of immunity
	\item Risks associated with the vaccination
	\item Vaccine efficacy
\end{itemize}
\vspace{0.5em}        
Length of immunity could be broadly categorized as `life-long' or `temporary'. Life-long immunity by vaccination can be achieved for some diseases, such as measles, mumps and rubella, while immunity obtained by vaccination for other diseases, such as influenza, would  be partial and/or only last until the end of the epidemic season. The length of immunity affects how frequently the vaccination game is played. If the vaccination offers life-long immunity,  an individual only gets to play the game once in his/her life, whilst if the vaccination confers temporary immunity, the individual gets to decide whether to get vaccinated every season, thus playing the game every season. Coverage of immunity defines the level of protection that a vaccine can provide. Full immunity may be granted for vaccines that are clinically proven to be fully effective and thus a vaccinated individual will be removed from the susceptible class and no longer exposed to the risk of infection. Where only partial immunity is granted (for example, in the case of influenza), a vaccinated individual is still placed in the susceptible class but with a reduced probability of infection.

Vaccination comes with some risks to individuals who get vaccinated, such as vaccine-induced morbidity or other side effects, though such a risk materializing is typically extremely unlikely. The risk of vaccination will nevertheless affect the expected payoff of vaccination, since this is a potential `cost' of vaccination \cite{static_group_ind}.

Vaccines, whether granting full or partial immunity, do not always meet their expected level of efficacy. Manufacturing defects have been considered in previous studies \cite{network_game_bauch}. As a result, a small percentage of vaccinated individuals do not receive any protection (full or partial), and should  be treated as unvaccinated individuals who are still prone to infection.

\paragraph{Social distancing}
Also known as behavioural change, social distancing generally refers to adoption of prophylactic measures to reduce physical contact between individuals, thereby minimizing the likelihood of infection spread.  Susceptible individuals who choose this strategy face some form of compromise, represented by extra cost, such as forfeiting on travel plans, or avoiding crowded environments in their daily lives. For those who do not adopt any change in lifestyle, the risk of infection remains relatively high but they are spared the extra cost related to social distancing. Individuals can switch between ``social distancing" or ``social non-distancing" depending on the payoff for each strategy by conducting cost-benefit assessments \cite{beh_imi3,beh_imi1,beh_imi2,reluga1}.

\subsubsection{Pay-off construction}
In addition, we consider different methods of payoff construction, without using this in our categorization explicitly. Assuming all players in the game are rational and their best interest is to obtain maximum gain, their decision-making relies purely on the outcome of cost-benefit assessments. They weigh up the payoff for each strategy, and at the end of each game, a net payoff is obtained. This is defined by a numerical value according to the strategy the individual adopts \cite{game_book} \cite{games_thomas}. In the epidemic context, there are several approaches to measure payoffs, such as:

\begin{itemize}
	\item Risk assessment: a qualitative measurement that often poses as a reflection on the risks of death from the infection \cite{static_bauch,onofrio_2011_imi_vac,samit_bauch_2011_delay,imi_bauch_bha,bauch_2005_imi}
	\item Monetary value: expected costs expressed as the true financial cost of choosing a certain strategy; this can be either a realistic quote or an estimate based on previous epidemic breakout, or a purely indicative index \cite{USA_Israel,mon_cost2,mon_cost,vac_influenza}
	\item Others: years to live, etc. \cite{QALY1,network_game_bauch,QALY3,QALY2}
\end{itemize}

A classic approach to constructing payoffs is to quantify risks, which can be specifically stated as morbidity (i.e., risk of death) or simply as the risk of infection as shown in \cite{static_bauch, static_group_ind}. In vaccination games, the risk of vaccination is often defined as the morbidity the vaccine induces; and the risk of non-vaccination is defined as the risk of infection, both simplified as a constant as seen in \cite{static_bauch, sto_book, sto_network_2010,sto_network_2012, sto_network_2012_memory,sto_network_2013}, and in equation \ref{const}:
\begin{equation}
\begin{cases}
E_{v}=-r_v \\
E_{i}=-r_i \\
r_v<r_i \\
\end{cases}
\label{const}
\end{equation} 
where $r_v$ represents the risk of vaccination and $r_i$  the risk of infection,with the constraint that the risk of infection is always greater than the risk of vaccination. \\

More sophisticated expressions for the risk of infection are not uncommon. These usually depend on the prevalence of infection, as seen in \cite{bauch_2005_imi, beh_imi1, beh_imi2, beh_imi3}. For example, in vaccination games, given that the payoff is measured by risks of morbidity, the payoff for non-vaccination increases  when the epidemic propagates and there are greater numbers of cases with infection. In this scenario, the payoff for not vaccinating is required to reflect the increasing risk of infection and thus defined as a function with variables relating to the size of the infected population \cite{bauch_2005_imi, beh_imi1, beh_imi2, beh_imi3}. Some studies also use a modified form to distinguish between the actual infection prevalence and the perceived prevalence, where the latter is meant to mimic the scenario in which individuals do not necessarily grasp the full extent of the current epidemic breakout. Perceived prevalence is thus built on current observation (force of infection $\lambda$) and memory of previous vaccination history \cite{sto_network_2012_memory, minority1, minority2}.

Some studies constructed payoffs by assigning monetary costs to each strategy \cite{USA_Israel,mon_cost2,mon_cost,vac_influenza}. In vaccination games, if individuals choose to vaccinate, they need to pay a price for the vaccine itself  and the related efforts are associated with a cost, financial and otherwise (e.g: time consumed). If individuals choose not to vaccinate, the cost may consist of the treatment cost and/or other consequences of getting infected (e.g., absence from work). For diseases with well-documented epidemic history and available vaccines, the financial cost of vaccination is calculated by estimating costs of vaccines and health-related utilities \cite{QALY3, USA_Israel}.  In social distancing games, the cost or payoff for each strategy (i.e., change behaviour or not) is less clear as this may include a myriad of factors and would vary significantly depending on individual circumstances. A common approach to solving such a problem is to use an indicative index value rather than the true financial cost in payoff construction so that the generality is maintained \cite{QALY1,QALY3,QALY2}. One popular measure is the quality-adjusted life year (QALY), which incorporates both the duration and magnitude of the effects on reduced health on patients to measure the disease burden \cite{QALY1}. QALY is calculated as follows:
\begin{equation}
\begin{aligned}
QALY &=Y \times U \\
0 \leq \ & U \leq 1 \\
\end{aligned}
\label{QALY}
\end{equation}
where $Y$ and $U$ represent \textit {years of life} and \textit {utility of life} respectively.\\ 

If one lives for a year in perfect health, one QALY is awarded (utility is equal to 1). A lower utility is assigned when one lives out an year in less than perfect health condition. Because the estimate of utility often comes from patient-based surveys, QALY remains an index for data-rich epidemics such as HPV \cite{QALY3} and measles \cite{QALY2}.   

When each individual's expected payoff is calculated, an aggregate pay-off for the entire community can be obtained, as shown in \cite{static_group_ind, USA_Israel, reluga1}. 

\paragraph{Epidemic-dependent payoff}
Often individuals refer to the current disease prevalence to assess their risk of infection; therefore the function of payoffs is often expressed using epidemic-dependent parameters. Information about the current disease prevalence can be relatively accurate and objective, assuming such information is timely and obtained from reliable sources, such as news published on trustworthy media channels or directly obtained from public health authorities \cite{beh_rev}.

Equation \ref{beh_game} shows an example of how payoffs are constructed using timely epidemic prevalence information in a social distancing game where the two strategies from which individuals can choose are to adopt altered behaviour (i.e., taking protective measures), $s_{alt}$, or live normally without any change, $s_{nor}$. By altering behaviour, it is assumed that one would stay away from highly infectious hotspots (e.g., highly populated areas) and consequently reduce the risk of infection \cite{beh_imi3}.
\begin{equation}
\begin{aligned}
E_{nor} & =-m_{nor}I(\tau) \\
E_{alt} & =-k-m_{alt}I(\tau)\\
\end{aligned}
\label{beh_game}
\end{equation}
where $m_{nor}$ and $m_{alt}$ are parameters that show the risk of symptoms development induced by strategies $s_{nor}$ and $s_{alt}$, and $m_{nor}> m_{alt}$. \\
Some recent models, however, claim that accurate information on current disease prevalence may not be readily available and/or accessible to all individuals due to information delay or limited sources of information \cite{samit_bauch_2010_imi,minority_2011}. An example would be information obtained by word of mouth within an enclosed community or neighbourhood. The resulting disease prevalence, known as the perceived disease prevalence, may not necessarily be a true reflection of actual disease prevalence,  and can be highly subjective \cite{sto_network_2012_memory}. Some models purposely calibrate true disease prevalence to a perceived case to mimic the inaccuracy of information encountered in reality. Equation \ref{payoff_perc} shows an example of payoff functions considering perceived disease prevalence. Payoffs are expressed in expected lifespan (number of years of life) after infection \cite{network_game_bauch}.  
\begin{equation}
\begin{aligned}
E_{inf} &=(1-\lambda_{perc})\alpha+\lambda_{perc}[(1-d_{inf})L] \\
E_{vac} &=(1-d_{vac})L \\
\lambda_{perc} & \approx \lambda \\
\end{aligned}
\label{payoff_perc}
\end{equation}
where $\lambda_{perc}$ is the perceived transmissiblity, obtained from the epidemic model,  $d_{inf}$ and $d_{vac}$ are deaths from infection and vaccination, respectively, $L$ represents `healthy' years of life with lifelong immunity and $\alpha$ is years to live with continued susceptibility.
\subsection{Taxonomy and trends in game-theoretic epidemic modelling}
\subsubsection{Taxonomy of studies}
The synthesis of game theory and epidemic modelling has evolved over the past decade. Starting from deterministic models combined with self-evaluating non-iterative population games, the focus eventually shifted to imitation-based and individual-based iterative games played by heterogeneously-mixed players. The shift reflects a desire to mimic more realistic  human behaviour, moving from homogeneous modelling that treats a population as a whole to heterogeneous individual-level networks that capture details. The game theoretic modelling of decision making  focuses  primarily on prevention measures, notably vaccination and behavioural changes. We argue that three criteria can be used to succinctly classify the existing literature:
\vspace{0.5em}

\begin{itemize}
	\item Type of population modelling
	\item Frequency of the game
	\item Type of strategy adoption
\end{itemize}   
\vspace{0.5em}
\begin{figure}[htb!]
	\graphicspath{{./fig/}}
	\includegraphics[scale=0.9]{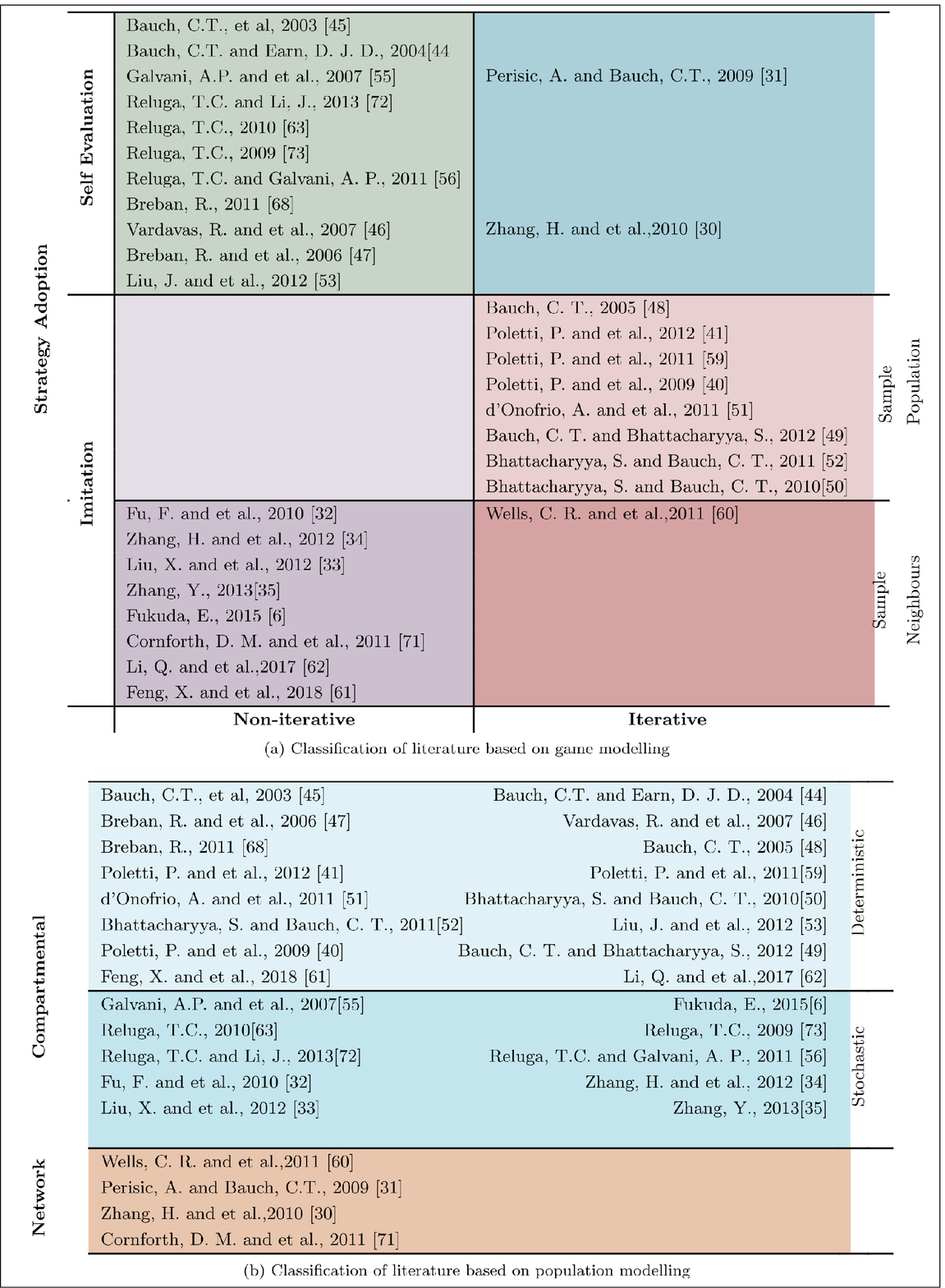}
	\caption{Taxonomy of present literature in  epidemic modelling with game theoretic decision making}. 
	\label{game_adop}
\end{figure}
Figure \ref{game_adop}  (a) and (b) shows that the investigations of vaccination decisions using different epidemic models and game settings have attracted considerable research interest. Earlier studies are clustered on the upper half of both sub-figures where the compartmental models with self-evaluation dominate. Imitation games gain more prominence in recent studies, especially when incorporated  with individual-based  models which typically have heterogeneous (contact network-based) mixing  of players.

`Type of population modelling' contrasts the population based (typically compartmental) and the individual network-based modelling approaches~\cite{onofrio_2011_imi_vac,static_bauch,bauch_2005_imi, imi_bauch_bha, samit_bauch_2011_delay, minority_2011, static_group_ind, mon_cost,USA_Israel,sto_network_2010, sto_network_2012_memory, network_game_b2,  sto_network_2012}. To re-iterate, population-based modelling refers to the modelling method where the population is represented as homogeneous, and properties of individuals such as their age, gender, travelling habits etc are not used in modelling the infectious disease dynamics. Thus, compartmental models such as SIS and SIR are sufficient for this modelling approach.  On the other hand, in network-based modelling, the population is assumed to be heterogeneous in terms of their attributes, and such attributes are explicitly used in modelling the spread of disease~\cite{onofrio_2011_imi_vac,static_bauch,bauch_2005_imi, imi_bauch_bha, samit_bauch_2011_delay, minority_2011, static_group_ind, mon_cost,USA_Israel}.

Studies which employ compartmental modelling can be further classified into those which use deterministic compartmental modelling and those that use stochastic compartmental modelling~\cite{sto_book, math_epi}. 

`Frequency of the game' distinguishes between non-iterative or iterative games. When the decision making is done only once, or seasonally within a particular `window of opportunity', non-iterative games are used \cite{sto_network_2010, sto_network_2012_memory, network_game_b2, minority1}. On the other hand, when the decision making is a continuous process, and strategies can be dynamically changed during the spread of the epidemic, iterative games are used \cite{onofrio_2011_imi_vac, bauch_2005_imi, imi_bauch_bha, samit_bauch_2010_imi, beh_imi1, beh_imi2, beh_imi3}.

We note a sub-classification, differentiating between the following two methods used in non-iterative population games (typically with deterministic SIR).  While some studies \cite{static_bauch, static_group_ind} adopt an analytical derivation, others \cite{reluga1, reluga_galvani} opt for ``Markov decision processes", which are stochastic models of individual decision-making, in calculating the expected value conditional on the vaccination decision in any given health state \cite{markov}.

`Type of strategy adoption' separates  self-evaluation from  imitation as a means to update strategies used by individual players.  As noted earlier, self-evaluation is used when a player considers only their past strategies and payoffs in deciding the future strategies, whereas imitation  is employed when a player stochastically adopts the strategies of other successful players. Imitation can occur either at a global level, where a player considers all other members of the population and stochastically imitates the strategies of the players with better past payoff, or at a local level where a player stochastically imitates only their (successful) neighbours.

In the lower right quadrant of  Figure \ref{game_adop} (a), along with the upper half of Figure \ref{game_adop} (b), a large set of studies use imitation and iterative population games, employing the deterministic SIR modelling. The inclusion of imitation dynamics introduces social influences on individuals, and is seen as a better representation of the reality compared with self-centred, non-communicating individuals. The implementation of imitation dynamics also allows for an adjusted time scale for information spread, so that the imitation process can take place at a faster or slower rate than the epidemic spread \cite{beh_imi1, beh_imi2, beh_imi3}.

In network-based studies, clustered in the lower half of Figure \ref{game_adop} (b),   the game involves self-centred individuals weighing up two strategies, by referring to memories and pursuing personal interest with no influence from the population.  Studies conducted in this setting investigate the ability of voluntary vaccination to prevent influenza epidemics \cite{minority1}, the connection between incentive/penalty and the vaccination coverage \cite{minority2}, and the impact of health newscasts publicizing disease incidence and/or vaccination coverage on epidemic dynamics\cite{minority_2011}.  The incorporation of contact networks is a more realistic representation of the social influences exerted on individuals, particularly with respect to behavioural change. This is because social actions take place within social locales defined by geographical, social, cultural and psychological factors, which can be represented by possible network links \cite{pattison2}. The implementation of network-based imitation is typically applied in voluntary vaccination for recurring epidemics (e.g., influenza).  Models in this area focus on studying the interrelations between epidemic quantifiers (vaccination coverage, infection prevalence, etc.)  and the vaccination cost by varying network structures~\cite{sto_network_2010, sto_network_2012_memory, network_game_b2,  sto_network_2012}.

Further refinements are added in later models to investigate the interplay between social influence and self-evaluation with respect to vaccination costs. At the population level, this involves committed vaccinators in the population \cite{sto_network_2012}, and, at the individual level, the memory-fading mechanism and risk evaluation \cite{sto_network_2012_memory}. In addition, a fitness factor may be considered to govern the likelihood of imitation, on top of the imitating behaviour in the network. The attempt to increase the vaccination coverage is made by modelling or simulating committed vaccinators, who would choose to vaccinate regardless of payoffs, and would act as ``role models" for their neighbours. If one is surrounded by vaccinators in the network, s/he is more likely to mimic the vaccination behaviour, contributing to vaccination coverage. The existence of committed vaccinators is realistic and  reduces the clustering of susceptible persons, ultimately extending the vaccination coverage. This observation is significant in a well-mixed contact network (well-mixed in terms of topology, not in terms of behaviour), where behavioural clusters are common: the committed vaccinators act as a strategy source and affect decision-making of their neighbours~ \cite{sto_network_2012}. 

Another branch of research combines the imitation dynamic with self-reflection, considering together the social influence from imitation and the self-centred evaluations resulting from memory fading and risk perception. For example, \cite{sto_network_2013} shows that imitation encourages vaccination behaviour at a low vaccination cost whilst the self-centred individuals are more likely to opt for vaccination if the vaccination cost is high. This  conclusion is demonstrated to  hold for random and scale-free networks. The incorporation of a memory-based mechanism indicated that the memory is very sensitive to vaccination cost. With low vaccination costs, the increase in memory promotes vaccination coverage; whereas when the vaccination cost is high, the increase in memory hinders vaccination coverage. 

Another point to note is that among the studies which directly use contact networks, one category of studies adopts different networks to model epidemic spread and individual contact networks \cite{network_static}, whilst others use the same network to model both epidemic transmission and contact pattern \cite{network_game_b2, network_game_bauch}.  For example, \cite{network_static} simulates epidemic spread on infinitely large random graphs where a connected pair of nodes bears a probability of transmission if one is infected. The dynamics of the vaccination game, in comparison, are built on individual's contact patterns across different networks: empirical networks in urban settings, or homogeneous networks and scale-free networks as model networks. Here, an individual's decision is finalized prior to the epidemic season (i.e., non-iterative) and remains a self-centred evaluation with some input from the topology of the network as the payoff assessment is dependent on the individual's degree $k$. 

Among the studies which use the same network to model  epidemic transmissions and contact patterns, \cite{network_game_bauch} employs a random network with a Poisson distribution to describe both the vaccination decision dynamics and the transmissibility (i.e., probability of infection) of a node. For a node, the magnitude of transmissibility is proportional to the number of infectious neighbours and is updated daily. The transmissibility is then used to compute the  payoffs for vaccination and non-vaccination. Thus, an individual assesses his/her payoff by referring to the number of infected neighbours (i.e., this depends on the topology of network), but the decision-making solely relies on his/her own evaluation and is not affected by the neighbour's decision.  Each individual's decision on vaccination contributes to the transmissibility function, as a vaccinated individual is removed from the susceptible class and will not be infected. This iterative feature bears significant resemblance to what has been seen in the synthesis between imitation dynamics and deterministic models \cite{bauch_2005_imi}, with different levels of social influence involved between self-evaluation and imitation. 

\subsection{Key findings}
In this section, we present the key findings of the considered studies.
\subsubsection{Key Finding 1}
\begin{figure}
	\graphicspath{{./fig/}}
	\includegraphics[width=0.9\columnwidth]{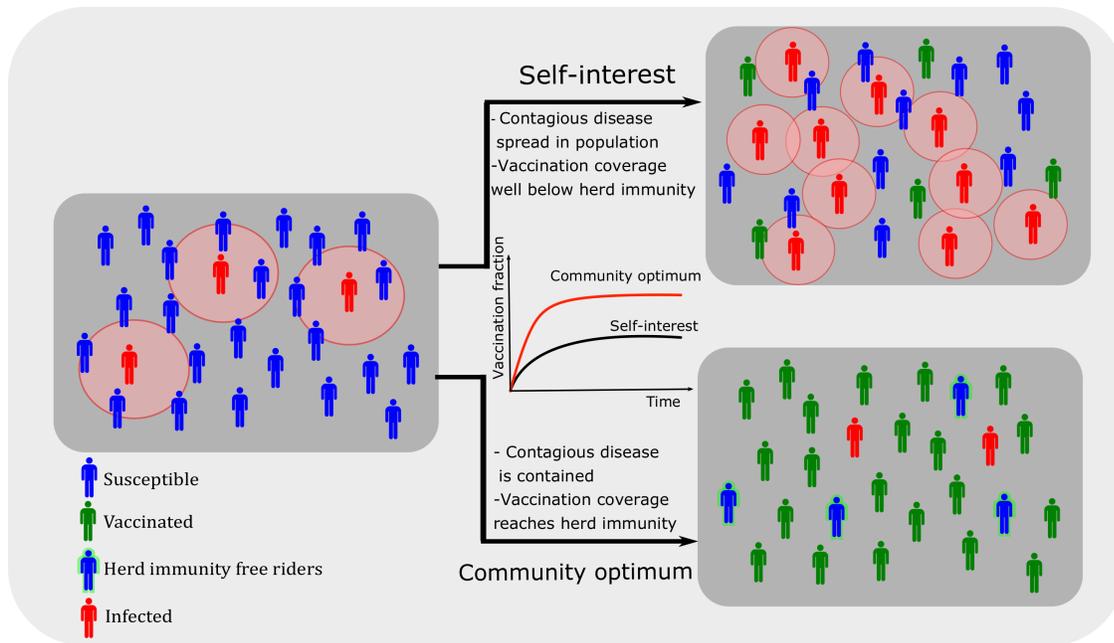}
	\caption{Key finding 1:  It is impossible to eradicate disease spread under voluntary vaccination schemes driven by the self interest of individuals: the level of vaccination needed to provide herd immunity to all non-vaccinated individuals, thereby eradicating the disease, is typically higher than the level of vaccination achieved by voluntary vaccination alone.}
	\label{key_1}
\end{figure}

\emph{Despite using different methods, all studies broadly agree, as shown in Figure \ref{key_1}, that \textit{a)} it is impossible to eradicate disease spread under voluntary vaccination, and \textit{b)} the vaccination level that is best for self-interest is always well-below the optimal level needed by the community.}

This is because unvaccinated individuals can always hope to escape the infection if the overall vaccination coverage is sufficiently high to curb the epidemic by reaching the ``\textit{herd immunity}" threshold.  Clearly, this is not attainable if each individual is purely selfish. These findings also hold with variants of the deterministic SIR models that divide the population in each health status into finer compartments by age and/or gender \cite{USA_Israel}.   The findings also indicate that the possibility of vaccination is dependent on the transmissibility of the disease, and is also proportional to the number of contacts the individuals have if they make decisions based on memories. 
\subsubsection{Key Finding 2}
\emph{The fraction of vaccinated population can (i) reach  zero (at low level of transmissibility below epidemic threshold), (ii) yield a wave-like vaccination behaviour (above a critical value of transmissibility), or (iii) converge to a stable value (either at low level of transmissibility above epidemic threshold, or at unrealistically high transmissibility), depending on the percentage of prevalence in the population and the value of $R_0$.}

\begin{figure*}[th]
	\centering
	\graphicspath{{./fig/}}
	\includegraphics[width=0.9\textwidth]{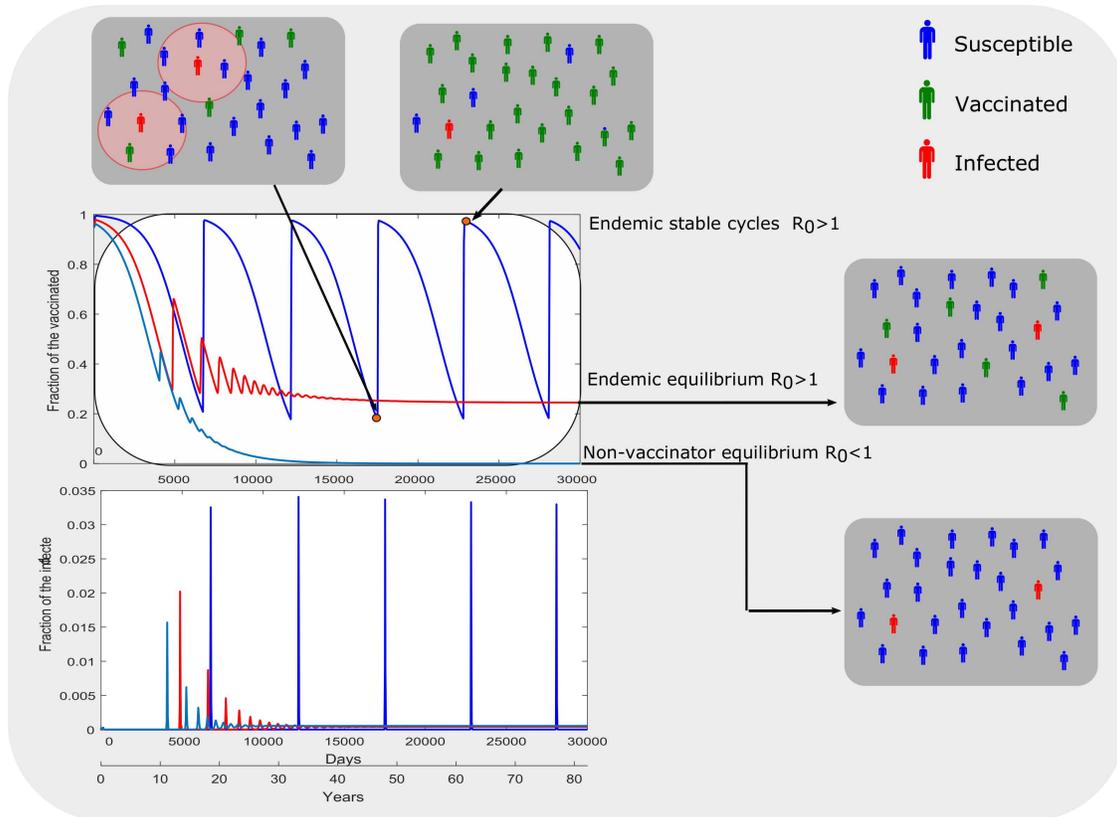}
	\caption{Key finding 2: The fraction of the people undertaking vaccination or social distancing may converge into stable values or display inter-dependent oscillatory behaviour with the prevalence percentage, depending on the value of $R_0$.}
	\label{key_2}
\end{figure*}

Some studies  \cite{bauch_2005_imi, beh_imi3} highlighted that the percentage of players who choose to intervene (by either taking vaccination or doing some form of  social distancing), and the incidence/prevalence percentage in the population, display inter-dependent oscillatory behaviour. In vaccination games, when vaccination coverage is high, the disease prevalence is depressed; subsequently, people have less incentive to vaccinate, and the prevalence percentage quickly recovers when the vaccination coverage drops \cite{bauch_2005_imi}. Similarly, if social distancing is adopted, if the number of susceptible individuals is not adequate to sustain the epidemic,  the epidemic peak decreases below the threshold value; while if the reproduction number for people not adopting social distancing $R_{0}^{a}$, is still greater than 1, the epidemic then picks up, creating wave-like oscillations \cite{beh_imi3}.  This is illustrated in Figure \ref{key_2}. 

\subsubsection{Key Finding 3}
\emph{The public health programs that offer incentives to vaccinate to family unit are less effective than the public health programs that offer incentives to individuals to vaccinate.}

\begin{figure}[th]
	\centering
	\graphicspath{{./fig/}}
	\includegraphics[width=0.9\columnwidth]{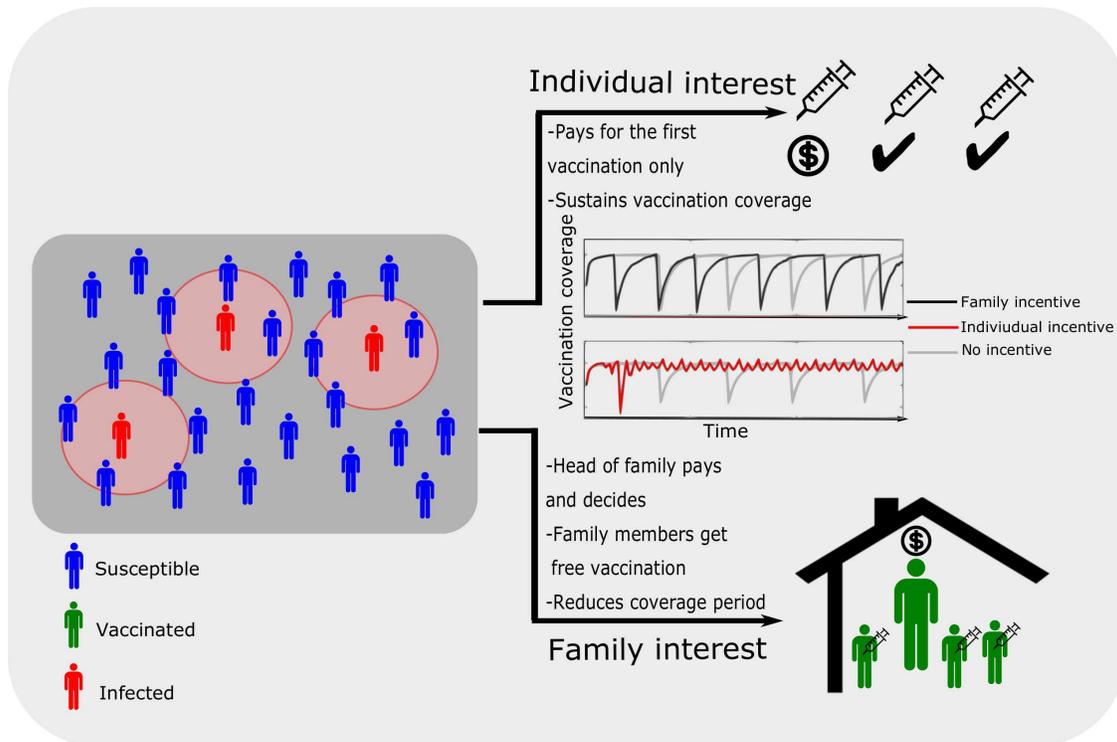}
	\caption{Key finding 3: free vaccination schemes are more effective if individuals are allowed to decide based on their self-interest. If the head of the family decides on behalf of the family, such free vaccination programs are less effective.}
	\label{key_3}
\end{figure}

As illustrated in  Figure \ref{key_3}, if free subsequent vaccinations are given to individuals who pay for their first vaccination, the critical vaccination coverage converges more rapidly and the threshold for epidemic spread is not reached. However, if the population is divided into families and the head of the family pays for his/her vaccination while the family members receive free vaccinations, the epidemic is more likely to develop. The reason for this is that such incentive programs reduce the number of players in the population and the head of family makes the decision on behalf of his/her entire family unit. Such  vaccination programs have been found to increase the frequency of influenza epidemics and reduce the duration of coverage \cite{minority2, minority1}. If no incentives are offered, the availability of epidemiological information plays a role in suppressing the epidemic severity. Assuming that (a) each individual is aware of an epidemic breakout, and (b) the media broadcasts information about the incidence and the vaccination coverage, or both, it was found that providing more epidemiological information does not necessarily improve the vaccination coverage, compared to broadcasting only the minimal information. As individuals become more knowledgeable about epidemiology, they could either become more attracted to vaccination, or more able to rely on herd immunity --- if they perceive that the current vaccination coverage is adequately high \cite{minority_2011}.

\subsubsection{Key Finding 4} 
\emph{Contact network topology plays an important role in affecting the level of vaccination coverage within a community.}

\begin{figure}[th]
	\centering
	\graphicspath{{./fig/}}
	\includegraphics[width=0.9\columnwidth]{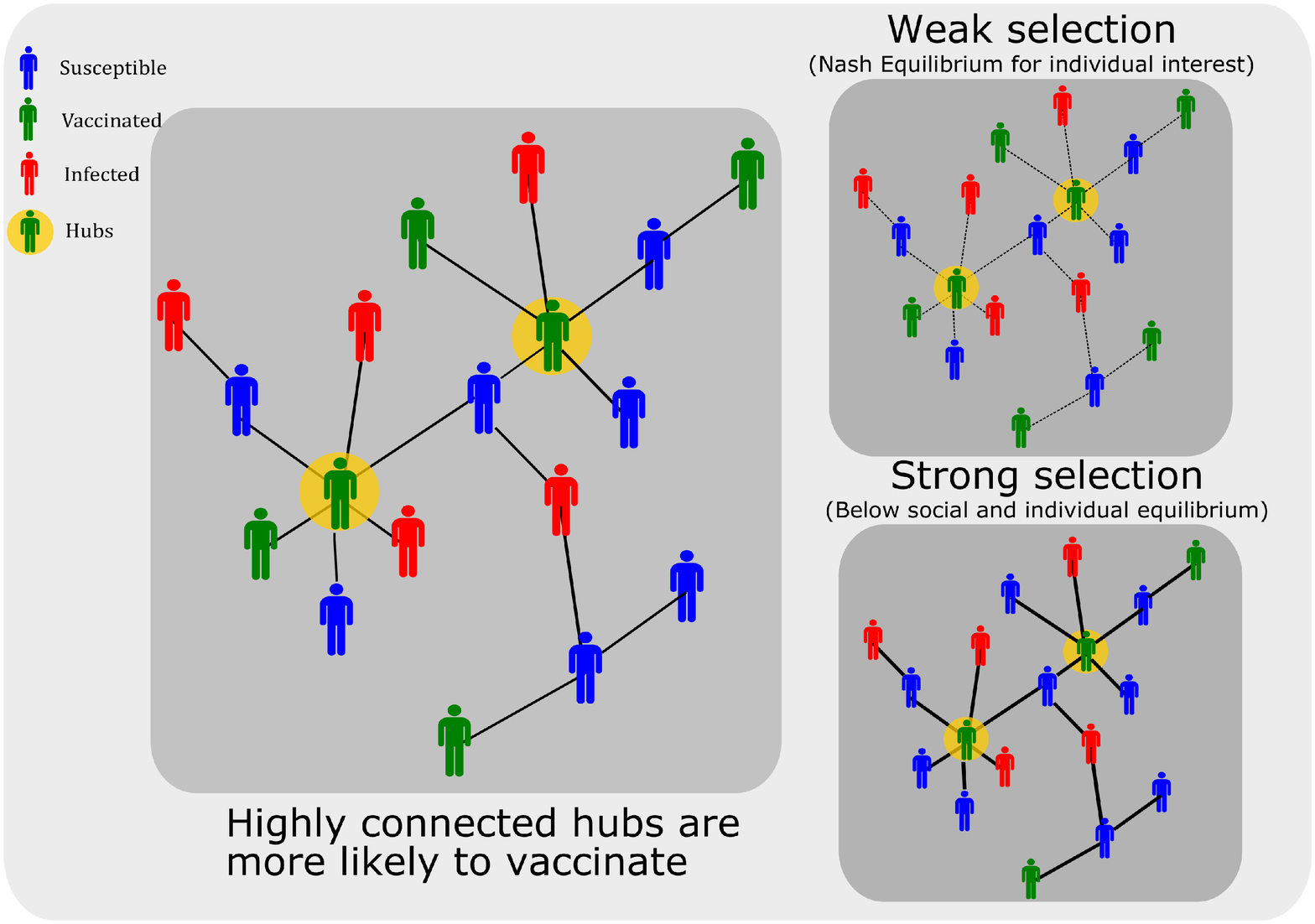}
	\caption{Key finding 4: Contact network topology plays an important role in affecting the level of vaccination coverage within a community.  Highly connected people are more likely to get vaccinated. }
	\label{key_4}
\end{figure}

Notably, the model proposed by \cite{sto_network_2010}  found that highly-connected individuals (network hubs) are more likely to be vaccinated because of the significant risk of infection. As illustrated in Figure \ref{key_4}, the strength of selection (the responsiveness to payoff difference) also affects the vaccination coverage to a great extent. With strong selection  and incomplete information, the vaccination coverage is well below both the social optimum and the Nash equilibrium (for individual's interest) because rational individuals are drawn to the success of free-riders after a single observation. Weak selection, by contrast, converges to Nash equilibrium in a well-mixed population. This is generally the case with over-exploitation of herd immunity, which in return delivers higher disease prevalence and lower vaccination coverage. Different topological structures are studied to investigate how topology affects the epidemic quantifiers \cite{sto_network_2010, sto_network_2012_memory, network_game_b2, sto_network_2012}. It is found that contact network topology can influence voluntary vaccination coverage and herd immunity but this effect is sensitive to the increase in the vaccination cost. A cost threshold exists beyond which small topological changes cause a significant reduction in the vaccination coverage and a rise in the infected cases, compared to homogeneous populations. This threshold-related effect is observed in both random and scale-free networks \cite{sto_network_2010}.  

\subsubsection{Key Finding 5}
\emph{Risk perception and vaccination costs are key factors in determining the level of vaccination within a community.}

\begin{figure}[th]
	\centering
	\graphicspath{{./fig/}}
	\includegraphics[width=0.9\columnwidth]{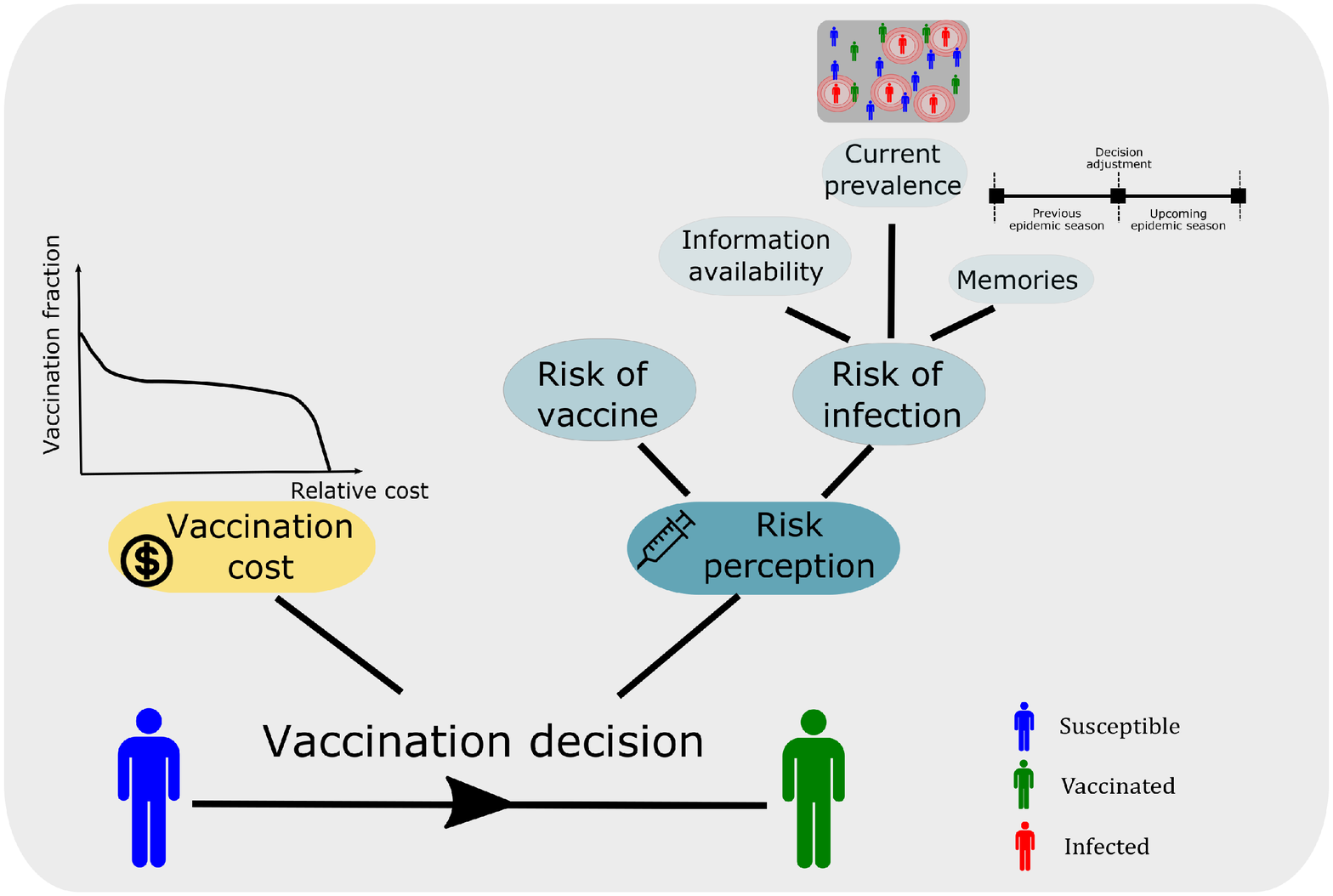}
	\caption{Key finding 5: Vaccination cost, risk of vaccine, and the perceived risk of infection are some factors influencing the vaccination decision.}
	\label{key_5}
\end{figure}

As illustrated in Figure \ref{key_5}, many factors influence the proportion of people in a population which decide to vaccinate voluntarily. Among these factors, risk perception and cost of vaccines (including monetary incentives)  have been  investigated  most extensively. The individual risk perception exerts a significant influence on epidemic spread. Two typical parameters, the risk of vaccine, and the risk of infection, were used in payoff assessment in early studies. Later, the assessment of risk perception  included more parameters representing different scenarios, such as memories of previous vaccination payoffs, individual risk threshold and the availability of information. The vaccination cost also heavily influences the vaccination decision. A low vaccination cost generally encourages the vaccination uptake until it reaches an intermediate range where individual decisions become insensitive to the cost change. A high cost threshold exists above which nobody would choose to vaccinate \cite{sto_network_2012_memory, sto_network_2010}.\\

The main finding from the models incorporating the risk perception resulting from observing a neighbour, was that a higher perception of risk encourages more vaccination. 
Interestingly, however, correlating the risk evaluation model with vaccination cost showed that for a small vaccination cost, the epidemic prevalence for the risk evaluation model is higher, compared to the original model without risk evaluation. The explanation  is that  the evaluation process creates a threshold in risk assessment,  and individuals with a high risk threshold have a low possibility of getting vaccinated.  The vaccination behaviour of the individuals in such models is not driven by the vaccination cost as strongly as in the models without risk evaluation. This suggests, firstly, that a lower vaccination cost does not necessarily suppress the epidemic level and/or encourage more vaccination, and secondly, that an individual's decision-making produces distinct outcomes if different conditions are applied \cite{sto_network_2012_memory}. \\

The behaviour-prevalence dynamics in populations modelled as random networks indicate that the epidemic spread is very sensitive to a change  of risk assessment made by the individuals facing a vaccination decision. The overall impact is exhibited on the population level: for example, in the final epidemic size and number of vaccinated persons. Another interesting finding is that voluntary vaccinations can contribute to disease eradication when individuals act for their own interest under certain conditions, contradicting the common conclusion that voluntary vaccinations cannot stop an epidemic outbreak. Furthermore, \cite{network_game_b2} builds on this behaviour-prevalence system by drawing comparisons with a scale-free network. Findings show that the scale-free network topology suppresses the epidemics more compared to  random networks, as the hubs tend to vaccinate,  and their high connectivity encourages the higher overall vaccination coverage.

\subsubsection{Summary}
Though using different modelling approaches, the existing literature appears to arrive at the consensus that human behaviour and decision making at the individual level have a significant impact on epidemic spread at the population level. Two such behaviours, social distancing and vaccination, have been widely studied. It has been found that voluntary vaccination decision making can curb the disease prevalence and reach Nash equilibrium  under some conditions, but that such an equilibrium is not able to eradicate diseases nor achieve herd immunity for the entire community. 

\section{Discussion: The emerging trends in game-theoretic modelling of interventions}
\begin{figure*}[t!]
	\centering
	\graphicspath{{./fig/}}
	\includegraphics[scale=0.95]{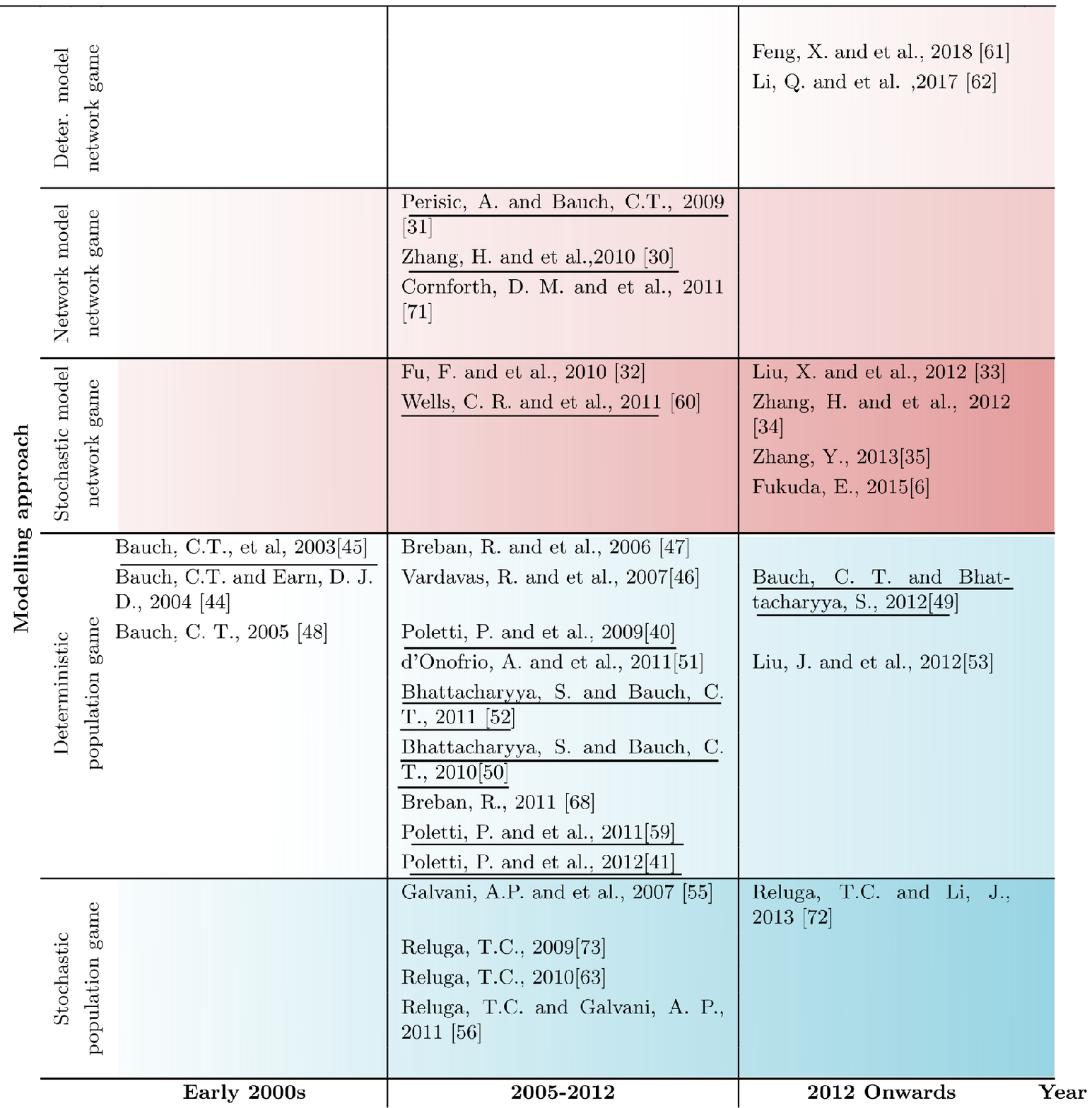}
	\caption{Research trends in game-theoretic decision making in infectious disease dynamics. Where the game used is non-iterative, the author names are not underlined: where the game used is iterative, the author names are underlined.}. 
	\label{res_time}
\end{figure*}

In preceding section, we presented  a possible taxonomy of literature that focuses on game theory and epidemic modelling. Here we consider the temporal evolution of the research studies, and observe the emerging trends. The relations between individual decision making and overall epidemic spread began to be explored in the early 2000s, and producing several directions in epidemic modelling methods and game-theoretic decision making models.  Figure \ref{res_time} shows the distribution of the papers covered in this review over time. It can be seen that the modelling efforts in the early stages focused on the use of deterministic models with self-centred game settings. This combination dominated early investigations, which generally only examined mass human behaviour. These earlier studies were built on the assumption of homogeneous mixing in the population, an assumption that quickly fell out of favour due to its inability to capture nuances in an individual's social life. Later on, still using deterministic models, attempts were made to divide the population into finer compartments based on demographic characteristics. More sophisticated models quickly emerged, incorporating networks into the epidemic modelling  to achieve heterogeneity. It was also found that the topology of networks plays an important role in controlling epidemic prevalence, so the search for a realistic representation of networks in the population continued. In this regard, the scale-free network model is, at present,  favoured as the most realistic,  while research efforts continue to develop model networks with topological properties that best mimic real-world contact networks. Stochastic models have also gained in popularity in recent years, aimed to model epidemic spreads on diverse contact networks. 

Turning our attention to game-theoretic models of decision making, we note that non-iterative games with self-evaluation  were replaced by iterative, imitation-based games. This allowed researchers to include  flexible windows of opportunity in decision-making, as well as social influences on all individuals within a population. Non-iterative population games with a simple self-evaluation of payoff assessment were widely studied in the early years. These represented the ideal case in which every individual performs an identical evaluation process while being perfectly rational and informed about current epidemic prevalence, however lacking memories or previous experiences. Later studies  distinguished between the perceived and actual epidemic prevalence. In making intervention decisions for an upcoming epidemic season, these studies model the individuals which consider not only the current payoff and/or risk assessment, but also the  `weighted memories' of previous vaccination experiences, including the memories of strategy adoption in previous seasons, and the memories of payoff.  
For instance, the adaptability parameter of memory, $\epsilon$ ($0<\epsilon<1$), controls how memories of strategy affect the strategy adoption in the next season, where $\epsilon=0$  means individuals will take the same strategy as in previous years, and $\epsilon=1$ means the strategy adoption is solely based on current payoff assessment, independent from previous experiences \cite{minority1}. 

Another set of studies using the imitation dynamics quickly attracted interest by connecting decision-making to a learning process of not only individuals, but the whole population. Consistent with the trend observed in the epidemic modelling method, the imitation games studied in later years focus on the individual level by limiting the learning process to people who have contacts with each other. Network-based modelling constrains individuals to play the imitation games only with their connected neighbours. The significance of the topology of contact networks has been studied extensively. Some studies \cite{sto_network_2012_memory} incorporated both self-evaluation and imitating contacts in the payoff assessment. While imitating neighbours, individuals also use memories of payoff, from previous vaccination experiences, to evaluate the total payoff for the current vaccination decision. 
For instance, the total payoff for a strategy may combine the current payoff and the historical payoffs using the  `weight of memory' parameter $w$  $(0<w<1)$. While $w=0$ means that the previous vaccination experiences are omitted, $w=1$ means the historical payoffs are given equal weight to the  current payoff.  In other words, the payoff functions include the parameters governing self-decision-making (e.g., memory, perception of risk, etc.) and those reflecting learning processes (e.g., strength of selection). 

\section{Conclusion}
We reviewed game-theoretic studies modelling both (a) a spread of contagion, and (b) the decision making of individuals, contemplating a number of intervention options to protect themselves from the epidemic spread. 

The systematic review, based on PRISMA search process, was followed by a succinct categorization of the research works, allowing us to summarize key findings and identify research trends.  

Specifically, the classification distinguishes (a) the type of population model, (b) the frequency of the games played, and (c) the type of strategy adoption. The `type of population model' contrasts between compartmental and network-based modelling. The compartmental models such as SIR (Susceptible-Infected-Recovered), and SIS (Susceptible-Infected-Susceptible) are further divided into deterministic and stochastic models. They focus on the aggregate behaviour of sections (compartments) of the population, whereas the  individual-based network models focus on individual behaviour. 

The `frequency of game' separates non-iterative and iterative games. Non-iterative games are used where the decision can be made only once in the lifetime of an individual, or only once during a season: for example, deciding to have the triple-vaccine administered must be done at a certain age, confers a life-long immunity, and thus cannot be reversed. 

Similarly, deciding to have a flu-vaccine is a decision which, while can be made multiple times within a person's life time, nevertheless must be made within a particular time-window before every flu season, and cannot be reversed once the flu season begins. Both these scenarios lend themselves to non-iterative games.  
On the other hand, iterative games are used to model decisions which can be changed during the course of an infection-spread. Most decisions regarding social distancing, for example, can be changed during the course of the epidemic, and thus such decisions can be modelled using iterative games.  
The `type of strategy adoption' differentiates between self evaluation and imitation.  An individual who uses self-evaluation only considers their past strategies and their pay-offs in making decisions,  whereas an individual adopting imitation stochastically copies the strategies of other successful members of the population. If a compartmental model is used, such choice imitates (the average of) the entire population, whereas when a network-based model is used with heterogeneous mixing, only the successful neighbours of a player would be imitated. 

The resultant taxonomy highlights  that the  choice of model depends on many factors such as the type of immunity the disease confers (if any), the type of immunity the vaccine confers (seasonal or lifelong),  the size of the population considered, and the level  of mixing in the population, among others.

Furthermore, we found that several studies used memory of previous experiences, including memory of previous strategies, as well as memory of previous payoffs, in both self-evaluation and imitation. In other words, players have been modelled to use both their own memories and the memories from the rest of the population (either memories of only the player's neighbours, or  the entire population). Often, studies used weighting mechanisms to limit the influence of memory in comparison to contemporary payoffs. 

Finally, we considered the current trends within the research field. We have shown that  early studies used compartmental models, whereas more recent studies prefer network-based models. The reason for this is that the network-based models are more realistic in scenarios where global information about the epidemic spread is limited and is not available in time, thus forcing players to depend on their own interactions to make decisions. Nevertheless, compartmental models are still widely used, possibly due to the fact that they are relatively easily solvable.  In terms of game frequency, however, we have observed that  non-iterative and iterative games both continue to be used, depending on the nature of the time window available to make intervention decisions. 

We believe that the presented review reduces a gap in literature, providing a useful summary of key findings and research trends in an expanding area of computational epidemiology. 

\section{Appendix} \label{PRISMA}
\begin{figure}[htb!]
	\graphicspath{{./fig/}}
	\includegraphics[width=0.8 \textwidth,keepaspectratio]{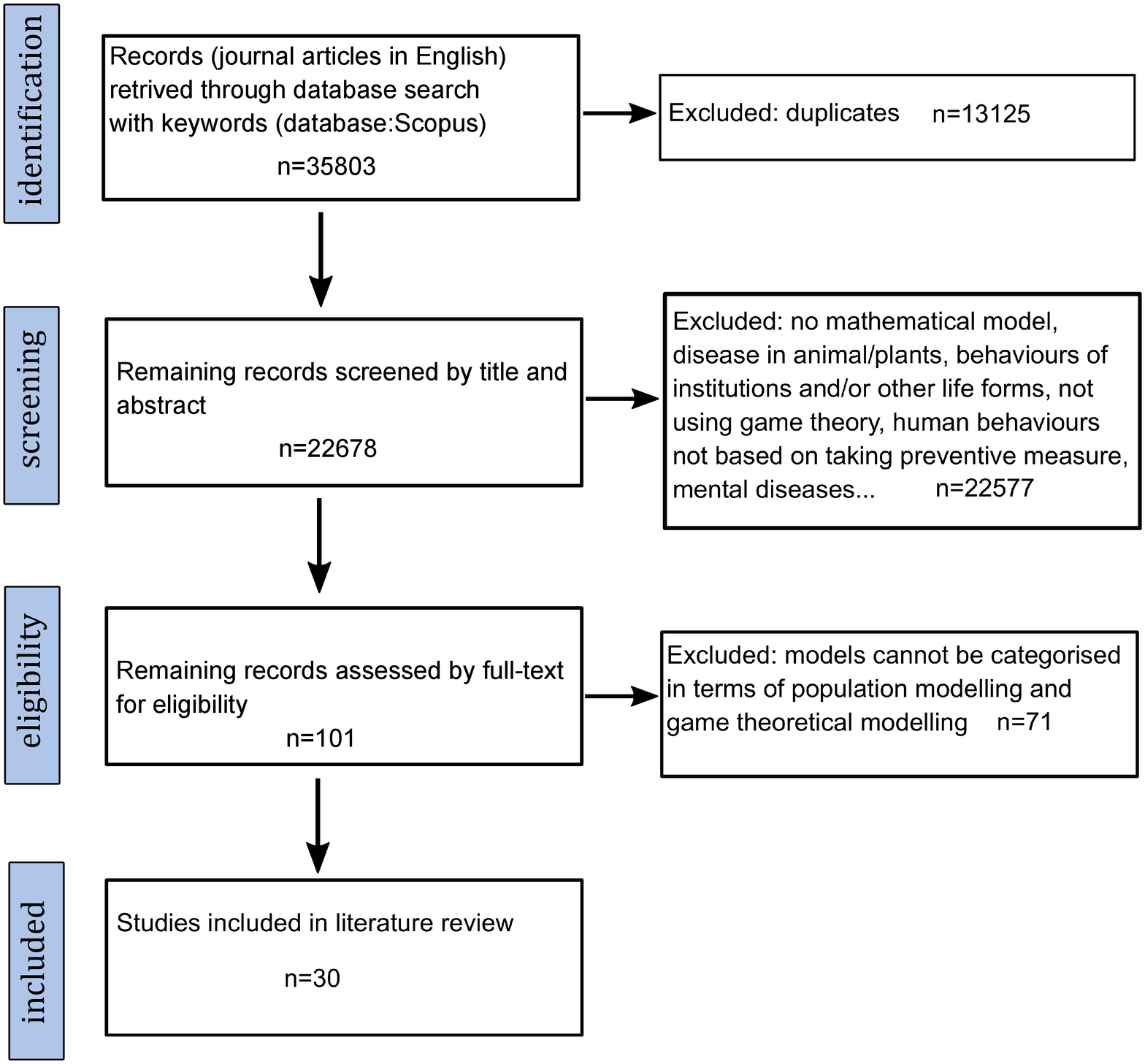}
	\caption{PRISMA flow chart of systematic search process}
\end{figure}

\subsection*{Declaration of interest}
The authors declare that they have no competing interests.

\bibliographystyle{unsrt}

\end{document}